\documentstyle[aps,pra,twocolumn]{revtex}
\begin{document}
\input{epsf}
\draft
\title{Nonlocality of Einstein-Podolsky-Rosen State
in Wigner Representation}
\author{Konrad Banaszek and Krzysztof W\'{o}dkiewicz\cite{unm}}
\address{Instytut Fizyki Teoretycznej, Uniwersytet Warszawski,
Ho\.{z}a~69, PL--00--681~Warszawa, Poland}
\date{\today}
\maketitle

\begin{abstract}
We demonstrate that the Wigner function of the Einstein-Podolsky-Rosen
state, though positive definite, provides a direct evidence of the
nonlocal character of this state.  The proof is based on an observation
that the Wigner function describes correlations in the joint
measurement of the phase space displaced parity operator.
\end{abstract}

Einstein Podolsky and Rosen (EPR)  in their argument about completeness of
quantum mechanics used  the following  wave function for a  system
composed of two particles\cite{epr}:
\begin{equation}
\label{eprfunction}
\Psi (x_{1},x_{2}) = \int_{-\infty}^{\infty}
e^{(2\pi i/h)(x_{1}-x_{2}+x_{0})p} \, \text{d}p.
\end{equation}
Despite its obvious simplicity, this wave function has not been explicitly
used in arguments relating the nonlocality of quantum mechanics with the
Bell inequalities. Following Bohm \cite{bohm} the EPR correlations have been
analyzed with the help of a singlet state of two spin-1/2 particles. For
this state the nonlocality of quantum correlations has been demonstrated
\cite{bell}.

Quantum correlations for position-momentum variables can be analyzed in
phase space using the Wigner distribution function. Using this
phase space approach to the EPR correlations Bell argued
\cite{bellwigner}, that the original EPR wave function
(\ref{eprfunction}) will not exhibit nonlocal effects because its joint
Wigner distribution function $W(x_{1},p_{1};x_{2},p_{2})$ is positive
everywhere and as such will allow for a local  hidden variable
description of momentum sign correlations.  

In local hidden variable theories
these and analogous correlations can be written in a form of a
statistical ensemble of two local realities
$\sigma({\bf a},\lambda_{1})=\pm 1$ 
and $\sigma({\bf b},\lambda_{2})=\pm 1$, for two
spatially separated detectors with certain settings labeled by 
${\bf a}$ and ${\bf b}$:
\begin{equation}
\label{lhv}
E({\bf a};{\bf b}) = \int \text{d}\lambda_{1} \int
\text{d}\lambda_{2}\, \sigma({\bf a},\lambda_{1})
\, \sigma({\bf b},\lambda_{2})\,  W(\lambda_{1};\lambda_{2}).
\end{equation}
In this relation $W(\lambda_{1};\lambda_{2})$ is a local, positive and
normalized
distribution of hidden variables $\lambda_{1}$ and $\lambda_{2}$. 
In the Wigner representation, these variables can be associated
respectively with the phase space realities $(x_{1},p_{1})$ and
$(x_{2},p_{2})$.  Bell's argument against the nonlocality of the EPR
wave function (\ref{eprfunction}) goes as follows. If the Wigner
function of the system is positive everywhere it can be used to
construct a local hidden variable correlation in a form given by
(\ref{lhv}) and accordingly the Bell inequality is never  violated. In
order to emphasize this point Bell used a  nonpositive Wigner function
to show that the momentum sign correlation function will violate local
realism. These  examples indicated a relation between the locality
and the positivity of the phase space Wigner function.

The relation between the EPR correlations and the Wigner distribution
function has been addressed in several papers
\cite{cetto,kimble,leonhardt,johansen,cohen}. Although the singular
character of the wave function (\ref{eprfunction}) and the
corresponding unnormalized Wigner function has been criticized, the
main point of the Bell argument relating the positivity of the Wigner
function to the lack of nonlocality of such a state has not been
questioned.  It has been argued that the problem of normalization can
be simply solved by a ``smoothing'' procedure of the original wave
function (\ref{eprfunction}).

An example of such a ``smoothing'' procedure, with a clear application
to quantum optics, has been the use of a two-mode squeezed vacuum state
produced in a process of nondegenerate optical parametric amplification
(NOPA) \cite{reid}.  The NOPA state has been generated experimentally
\cite{kimble} and applied to discuss the implications of the positivity
of the phase space Wigner function  on the Bell inequality
\cite{leonhardt}.

These discussions have led to rather ambiguous results. On one hand, it
has been argued that the quantum description for the system of the NOPA
as well as for the system originally discussed by EPR is consistent
with deterministic realism \cite{kimble}. From this remark one can
conclude that the EPR wave function (\ref{eprfunction}) cannot be used
to test direct violations of the Bell inequality. This  rather vexing
conclusion indicates that tests of quantum nonlocality have to rely not
on the original EPR wave function but on Bohm's spin-1/2 system or on
exotic states described by negative Wigner functions. On the other
hand, attempts have been made to design an experiment which would
reveal the nonlocality of the EPR state \cite{cetto,cohen}.

The purpose of this Rapid Communication is to demonstrate that the positive
definite Wigner function of the EPR state provides a direct evidence of
the nonlocality exhibited by this state.  We shall show that the
positivity or the negativity of  the Wigner function has a rather weak
relation to the locality or the nonlocality of quantum correlations. In
fact we shall show that the NOPA wave function violates the Bell
inequality and that the original  EPR wave function (\ref{eprfunction})
exhibits strong nonlocality, but one should be careful with the
singular limit of strong squeezing (in this limit the NOPA state
reduces to the EPR state). The NOPA phase space will be parameterized
by two complex coherent states amplitudes $\alpha$ and $\beta$
corresponding respectively to   $(x_{1},p_{1})$ and $(x_{2},p_{2})$.

The starting point of our proof is an observation
that the two-mode Wigner function $W(\alpha ;\beta)$ can be expressed as
\begin{equation}
\label{Eq:Walphabeta}
W(\alpha ;\beta) = \frac{4}{\pi^2} \Pi(\alpha ;\beta)
\end{equation}
where ${\Pi}(\alpha,\beta)$ is a quantum expectation
value of a product of displaced parity operators:
\begin{equation}
\label{Eq:Pialphabeta}
\hat{\Pi}(\alpha ;\beta) =  \hat{D}_1(\alpha) (-1)^{\hat{n}_1}
\hat{D}^{\dagger}_1(\alpha) \otimes \hat{D}_2(\beta)
(-1)^{\hat{n}_2} \hat{D}_2^{\dagger}(\beta) .
\end{equation}
The connection of the parity operator $(-1)^{\hat n}$ with the Wigner
function provides an equivalent definition of the latter
\cite{wignerparity}, as well as a feasible quantum optical measurement
scheme \cite{wignerparityexp}. In the above formula,
$\hat{D}_1(\alpha)$ and $\hat{D}_2(\beta)$ denote the unitary phase
space displacement operators for the subsystems $1$ and $2$.

As the measurement of the parity operator yields only one of two
values: $+1$ or $-1$, there exists an apparent analogy between the
measurement of the parity operator and of the spin-1/2 projectors. The solid angle
defining the direction of the spin measurement is now replaced by the
coherent displacement describing the shift in phase space. Consequently,
all types of Bell's inequalities derived for a correlated pair
of spin-1/2 particles can be immediately used to test the nonlocality
of the NOPA wave function. The two NOPA field modes are equivalent to an
entangled state of two harmonic oscillators. As 
Eq.~(\ref{Eq:Pialphabeta}) clearly
demonstrates, the correlation functions measured in such experiments
are given, up to a multiplicative constant, by the joint Wigner function
of the system. As a consequence we have the fundamental relation:
\begin{equation}
E({\bf a};{\bf b})\equiv \Pi(\alpha ; \beta).
\end{equation}

The original EPR state (\ref{eprfunction}) is an unnormalizable delta
function. In order to avoid problems arising from this singularity, we will
consider a normalizable state that can be generated in a NOPA. Such a state
is characterized by the dimensionless effective interaction time $r$ (the
squeezing parameter). The Wigner function of this NOPA state is well
know\cite{kimble,leonhardt} and is given by
\begin{eqnarray}
\Pi(\alpha;\beta)
& = & 
\exp[-2 \cosh 2r (|\alpha|^2 + |\beta|^2) \nonumber \\
& & + 2 \sinh 2r (\alpha
\beta + \alpha^\ast \beta^\ast)].
\end{eqnarray}
The Wigner function of the original EPR state (\ref{eprfunction}) is obtained
in the limit $r\rightarrow \infty$.

The correlation function is measured for any of four combinations
of $\alpha=0,\sqrt{\cal J}$ and $\beta=0,-\sqrt{\cal J}$, where
${\cal J}$ is a positive constant characterizing the magnitude of the
displacement. From these quantities we construct the combination
\cite{CHSH}:
\begin{eqnarray}
\label{B}
{\cal B} & = & \Pi(0;0) + \Pi(\sqrt{\cal J};0)
+ \Pi(0;-\sqrt{\cal J}) - 
\Pi(\sqrt{\cal J}; -\sqrt{\cal J})
\nonumber \\
& = & 1 + 2 \exp(-2{\cal J} \cosh 2r) - \exp(-4{\cal J}e^{2r}),
\end{eqnarray}
which for local theories satisfies the inequality $-2\le {\cal B} \le
2$.  Let us note that one of the components of the above combination
desribes perfect correlations:  $\Pi(0,0) = 1$, obtained for a direct
measurement of the parity operator with no displacements applied.  This
is a manifestation of the fact that in the parametric process photons
are always generated in pairs.

\begin{figure}

\vspace{4cm}

\setlength{\unitlength}{0.675cm}
\centerline{%
\begin{picture}(0,0)
\put(0,0){\makebox(0,0)[c]{\epsfxsize=3.375in\epsffile{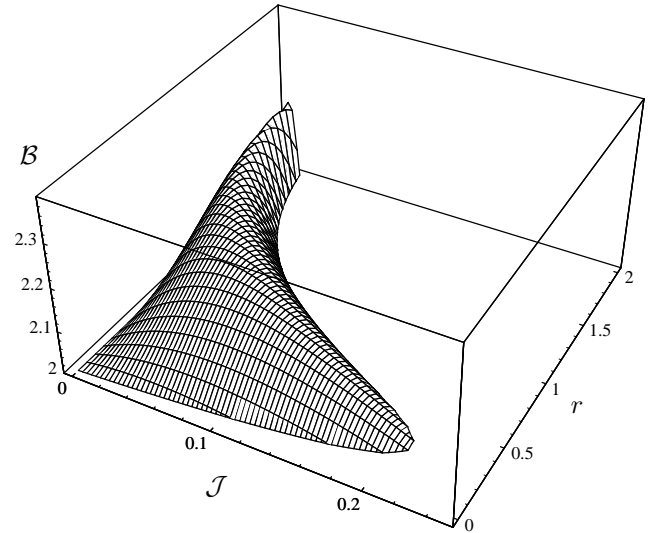}}}
\put(-2.6,-4.5){${\cal J}$}
\put(4.6,-2.9){$r$}
\put(-6.25,2.0){${\cal B}$}
\end{picture}}

\vspace{4cm}

\caption{Plot of the combination ${\cal B}$ defined in 
Eq.~(\protect\ref{B}). Only values exceeding the bound
imposed by local theories are shown.}
\label{Fig:B}
\end{figure}

As depicted in Fig.~\ref{Fig:B}, the result (\ref{B}) violates the
upper bound imposed by local theories. With increased $r$, the
violation of the Bell's inequality is observed for smaller ${\cal J}$.
We will therefore perform an asymptotic analysis for large $r$ and
${\cal J} \ll 1$. In this regime we may approximate $\cosh 2r$
appearing in the argument of the first exponent in Eq.~(\ref{B}) just
by $e^{2r}/2$. Then a straightforward calculation shows that the
maximum value of ${\cal B}$ (for this particular selection of coherent
displacements) is obtained for 
\begin{equation}
{\cal J} e^{2r} = \frac{1}{3} \ln 2,
\end{equation}
and equals ${\cal B} = 1 + 3 \cdot 2^{-4/3} \approx 2.19$. Thus, in the
limit $r\rightarrow \infty$, when the original EPR state is recovered,
a significant violation of Bell's inequality takes place. This result
has been obtained without  any serious attempt to find the maximum
violation (for this purpose one should consider a general quadruplet of
displacements).  Let us note that in order to observe the nonlocality
of the EPR state, very small displacements have to be applied,
decreasing like ${\cal J} \propto e^{-2r}$. This shows the subtleties
related to the original EPR state (\ref{eprfunction}) and the need for
considering its regularized version.

This discussion shows, that despite conflicting claims, the original
EPR wave function (\ref{eprfunction}) exhibits strong nonlocality. The
violation of the Bell inequality is achieved for a state that is
described by a positive  Wigner function. This example puts to rest
various conjectures, relating the positivity or  the negativity of the
Wigner function to the violation of local realism. We have shown that
in quantum mechanics, the correlation (\ref{lhv}) can be a Wigner
function itself. This is  due to the fact that the Wigner function can
be directly associated with the parity operator. This operator can be
measured in a photon-photon coincidence experiment.

Apparently, the Wigner representation cannot serve as a model
local hidden variable theory describing the joint parity measurement.
A straightforward explanation of this fact is given by expressing
the correlation function $\Pi(\alpha ; \beta)$ in the form
analogous to Eq.~(\ref{lhv}):
\begin{eqnarray}
\Pi(\alpha ; \beta)
& = &
\int \text{d}^2 \lambda_1 \int \text{d}^2 \lambda_2 \,
\frac{\pi}{2} \delta^{(2)}(\alpha - \lambda_1) \nonumber \\
& & \times
\frac{\pi}{2} \delta^{(2)}(\beta - \lambda_2)
W(\lambda_1 ; \lambda_2), 
\end{eqnarray}
where $\lambda_1$ and $\lambda_2$ are now complex phase space
look-alikes of hidden variables.
Though the outcome of the parity measurement may be only $+1$ or $-1$,
the analog of local realities appearing in the Wigner representation
is described by unbounded delta-functions
\begin{eqnarray}
\sigma({\bf a},\lambda_{1}) & \equiv & \frac{\pi}{2}
\delta^{(2)}(\alpha - \lambda_1), \nonumber\\
\sigma({\bf b},\lambda_{2}) & \equiv & \frac{\pi}{2}
\delta^{(2)}(\beta - \lambda_2), 
\end{eqnarray}
which makes the Bell inequality void.

A tempting aspect of the Wigner representation is the interpretation of
quantum mechanics in classical-like terms in phase space. One well
known difficulty with this approach is the negativity of the Wigner
function \cite{ghosts}. The example discussed in this Communication
shows, that quantum mechanics manifests its nature also in another,
equally important way: the Wigner representations of quantum
observables cannot be in general interpreted as phase space
distributions of possible experimental outcomes.  In particular, the
Wigner representation of the parity operator is not a bounded reality
corresponding to the dichotomic result of the measurement. This enables
violation of Bell's inequalities even for quantum states described by
positive definite Wigner functions.

{\it Acknowledgements.} This research was partially supported by
the Polish KBN grants and by Stypendium Krajowe dla M{\l}odych
Naukowc\'{o}w Fundacji na rzecz Nauki Polskiej.

\end{document}